# Domain ontology and multi-criteria analysis for enterprise modeling

Sabria Hadj Tayeb[1], Myriam Noureddine[2]

[1]**Preparatory school in Science and Technology of Oran (EPSTO), 31000/Oran, Algeria**
*s.hadjtayeb@gmail.com*

[2]**Computer Science Department, University of Science and Technology of Oran (USTO), 31000/Oran, Algeria**
*myriam.noureddine@univ-usto.dz*

**Abstract**

Knowing that an enterprise is a complex reality, it is necessary to develop a modeling framework allowing the description of system structure and dynamics that alter the structure. The concept of enterprise modeling addresses this need and many techniques have emerged. Our goal is to provide leaders of Algerian enterprise an overview of modeling techniques. Thus these managers may elect, in collaboration with the University, the modeling technique best suited to their requirements.

We believe that this could be a step towards an effective reorganization of the enterprise leading.

This article proposes a domain ontology and multi-criteria analysis in the frame of modeling enterprise. Our approach is based on two stages using the Protégé tool for the technique representation and the PROMETHEE method for their evaluation. The result is a ranking between the different techniques, which allows selecting the most appropriate methodology according to the criteria for a given enterprise.

## 1. Introduction

Faced with economic competition, enterprises are forced to improve their work techniques, their inner and external working. In this context, these enterprises must adopt a modeling framework to describe the system structure and dynamics that alter the structure over time.

Enterprises in Algeria are no exception to this rule, but we found on the ground when the Algerian enterprises want to adopt a new organization integrating the automation of their services, they don't have technical or in the best case, they use the MERISE method, well known in engineering information systems, but already old. In recent years, the concept of modeling has become a paramount concern for any enterprise. Many modeling techniques have then emerged, based on scientific concepts in the context of enhancing enterprise performance. Our goal is to provide leaders Algerian enterprises an overview of modeling techniques, with their characteristics. To achieve this goal, we developed a framework of knowledge through a list of criteria in order to observe the largest number of technical characteristics. This article proposes a domain ontology and multi-criteria analysis for the choice of technique in the modeling enterprise. We present in the second section the modeling techniques selected and the proposed approach is described in the third section. The fourth section presents a validation of the approach through a real case of an enterprise, followed by the conclusion.

## 2. Enterprise modeling

There are several definitions of enterprise modeling [5]. We accept that the enterprise modeling is the representation of the structure and operations of the enterprise to improve its performance. This vision concerns the modeling of information system of the enterprise centered on the production system.

2.1 Techniques

The modeling techniques are diverse; each is based on scientific concepts. Under the technical term [13], we group techniques (techniques for solving a single problem leading to a model), methodologies (all technologies) and reference architectures (context support a methodology).

We are interested in this article to a GRAI technique and its variants, as well as technical CIMOSA, PERA and GERAM. These techniques have been used for standardization [3, 11] through the construction standards of enterprise modeling (ENV 40003, ISO 15704 and EN / ISO 19439). We add the technical MERISE because it is used in Algerian enterprises.

2.2 Description of techniques

**MERISE:** The technique MERISE [14] provides both of process, models, formalisms and standards for the design and implementation of information systems enterprise.

**GRAI:** The technique GRAI [6] was developed in the laboratory GRAI3 of University of Bordeaux. It objective is the modeling aspects of the decision taken during the analysis phase or design enterprise.

**CIMOSA:** CIMOSA [16] is considered one of the modeling approaches that generated the most research work. Its purpose is to define precisely the objectives of the enterprise, manufacturing strategies and managing the system in an environment of perpetual change.

**PERA:** PERA [18] is a complete architecture of industrial engineering environments developed by the Purdue Laboratory of Applied Industrial Control in the U.S. It aims to design large systems.

**GERAM:** GERAM [2] is generalized reference architecture CIMOSA GRAI, PERA to represent the integrality of enterprise.

**GIM:** GIM [12] is a variant of GRAI methodology. It allows to modeling the existing system and designing the target system model from the analysis of existing and objectives assigned to the system.

## 3. Proposed approach

The approach of the choice of modeling technique is based on the meta-model given in Fig 1, following two steps:

- The knowledge representation techniques and characteristics (criteria) using a domain ontology, and realized with the environment Protégé.

- The multi-criteria analysis according the outranking method PROMETHEE (Preference Ranking Organization Method for Enrichment Evaluations).

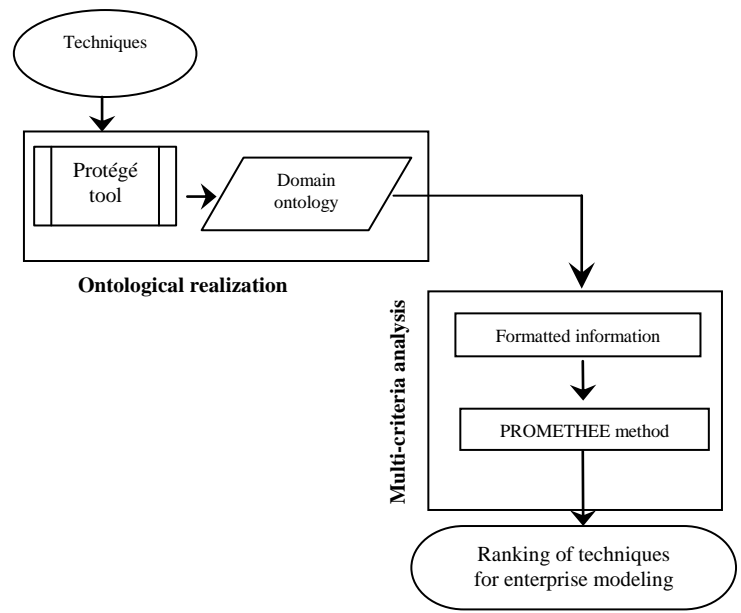

Fig.1 Meta-model of the Approach

The decision system was implemented following software that supports both steps.

3.1 Ontological representation of knowledge

3.1.1. Identification and hierarchy of criteria

From the study of modeling techniques and real cases of an enterprise, we have identified fourteen criteria [9].

1. Generic model, if the model is applicable to a wide type of enterprise.
2. Formalism, if the formalism is adaptable to corporate actors.
3. The cycle of life, captured by the technique of modeling enterprise.
4. Software support: this criterion represents the generation or not a support to facilitate the construction and operation of the model.
5. Learning criterion that reflects the mastery or technology enterprise modeling.
6. Ease of use, this criterion is the ability to assimilate technology enterprise modeling.
7. Time: this is necessary to describe the properties of states, their changes over time. It used also to represent the processes evolving in parallel and influence against each other.
8. Function and flow decisions; these two criteria are necessary for decision making.

10. Human resources; this criterion is chosen for the taking or not the human aspect. It is necessary to describe the skills, roles, responsibilities and knowledge of human actors in the production process.

11. Functional, organizational, resource and informational views, these four criteria are the types of views offered by technology.

We have grouped sequentially these criteria into five sets of criteria, giving a meta-family $F = \{f_1, f_2, f_3, f_4, f_5\}$

$f_1 = \{f_{11}, f_{12}, f_{13}, f_{14}\}$ lists the criteria for the model.
$f_2 = \{f_{21}, f_{22}, f_{23}\}$ is said general criterion.
$f_3 = \{f_{31}, f_{32}\}$ makes reference to the structure.
$f_4 = \{f_{41}\}$ contains the criteria for the resources.
$f_5 = \{f_{51}, f_{52}, f_{53}, f_{54}\}$ includes different views.

To represent the knowledge base consisting of modeling techniques and criteria, we choose to build the domain ontology.

### 3.1.2. Realization of the domain ontology

According to [7], ontology [8] is a formal and explicit specification of a conceptualization. To build the ontology, we adopted the protected environment Protégé.

Protégé version 3.1 [10] is a java tool free to use, it is produced and made available by the Stanford Medical Informatics laboratory.

The creation of the domain ontology is based on three stages [9].

**Creating concepts**: The group of criteria $F$ and the term 'technique' $T$ will be formalized by the creation of two concepts at the same level. The construction of the concept hierarchy will be done by creation of sub-concepts for any meta-criterion $f_i$ of the concept $F$. The following figure (Fig. 2) illustrates the hierarchy of concepts.

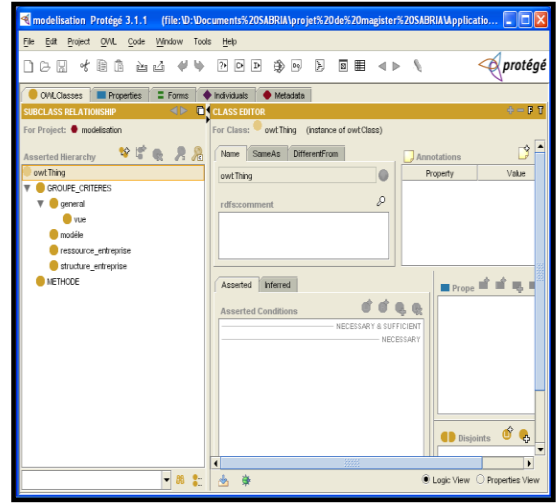

Fig. 2 Hierarchy of concepts

**Creating attributes:** This step involves the creation of attributes representing all the identified criteria.

**Creating forums**: The exploitation of ontology is made through the forums by giving values to any attribute on a technique.

For example, the figure 3 represents the instance on the technique CIMOSA following the structure enterprise criteria ($f_3$). The study of a technique has assigning values to following attributes:

Decision flow ($f_{31}$): unknown; decision function ($f_{32}$): partial.

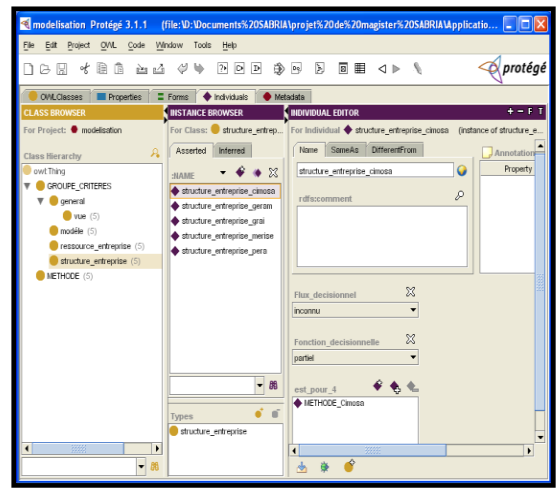

Fig. 3 Creating forums

The ontology obtained is translated by a schema graph (Fig. 4). This vision emphasizes the hierarchy of different elements and the relationships between them.

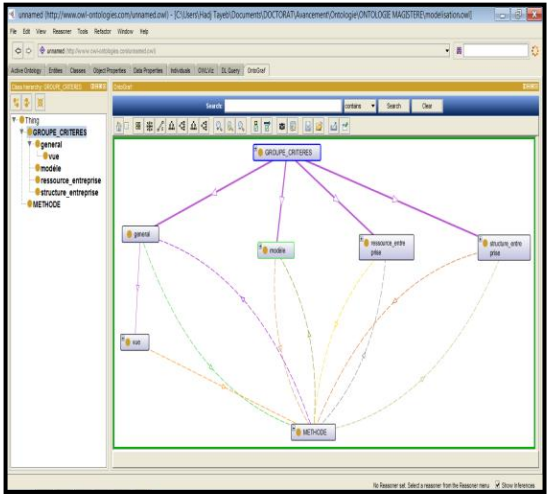

Fig. 4 Visualization of the ontology

## 3.2 Multi-criteria analysis by PROMETHEE II

The second phase of our support system for choosing a technique is based on a multi-criteria decision analysis.
A multi-criteria evaluation includes four sequential steps [1] of any approach to multi-criteria analysis. We described these steps in the following figure (Fig. 5).

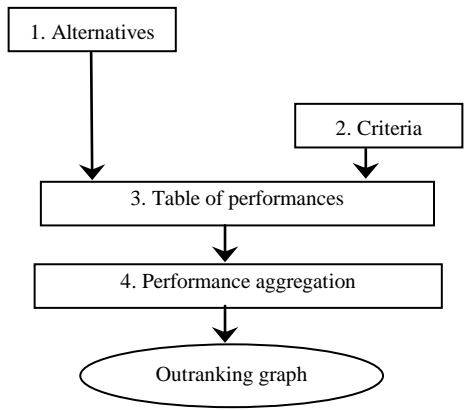

Fig. 5 Multi-criteria evaluation approach

### 3.2.1. The alternatives

The alternatives in our context are the different modeling techniques enterprise.

Initially, we have five alternatives; each of them represents a technique for modeling enterprise: MERISE, GRAI, CIMOSA, PERA and GERAM.
The alternatives are generic since we can add them using the tool Protégé.

### 3.2.2. The Criteria

The criteria used are the fourteen criteria identified above. Each criterion is evaluated for a technique enterprise modeling. Each $f_{ij}$ refers to the criterion $j$ belonging to family $i$ with its assigned values possible.

### 3.2.3. The table of performance

This table grouped the values of criteria for each alternative. We have developed a relevant scale from 0 to 4 representing the evaluation of all criteria according to the technical enterprise modeling.
Following this assessment, the performance table of criteria is established for the various alternatives to evaluate.

### 3.2.4. Performance aggregation

Among the techniques of multi-criteria analysis [1], we opt for an approach of partial aggregation of performance according to the outranking method PROMETHEE.

PROMETHEE [4, 15, 17] is to establish a process of numerical comparison of each action (technical enterprise modeling) compared to all other alternatives for a set of criteria. The result of this comparison allows the classification of alternatives ordered from best to worst.

- Characterization of criteria in PROMETHEE: We take a same weight and the first form for all criteria.
  The credibility matrix is calculated from the degree of preference $P_j(a_i, a_k)$ for any pair of alternatives for each criterion j.
  The calculation of the preference index $\Pi$ for any pair of alternatives $(a_i, a_k)$ is given by the formula 1:

$$\prod(a_i, a_k) = \sum_{j=1}^{n} P_j(a_i, a_k)/n \quad (1)$$

where n is the number of criteria.

- Outranking in PROMETHEE: There are two variants of the technique PROMETHEE and we opt for the version II.

PROMETHEE II deals with a complete ranking, while PROMETHEE I provides a partial ranking. The alternatives are classified according the order of qualification of each alternative along the net flow (formula 2). This net flow (ϕ) is defined from the positive flow (ϕ$^+$) and the negative flow (ϕ$^-$), defined respectively as the strength and the weakness of an alternative over the others.

$$\phi(a_i) = \phi^+(a_i) - \phi^-(a_i) \quad (2)$$

Generally, the final ranking is giving through a classical graph. To have more visibility, we propose the ranking in two forms of points or histogram of different generic alternatives.

## 4. Case study

The system validation decision has been made by implementing the technique PROMETHEE, then by its application on a case example of an enterprise.

### 4.1 Enterprise Description

We choose an enterprise [9] dedicated to the manufacture of glass bottles.

The manufacturing process (Fig.6) includes four steps:

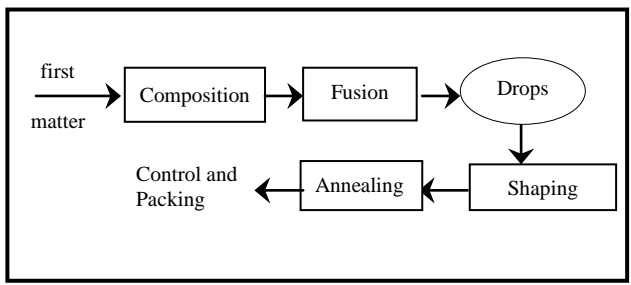

Fig. 6 Organization process of bottle manufacturing

- **Composition step**: It is a mixture of different components (sand, limestone…). Then, this mixture is fed into the furnace. This step is done automatically by computer-guided machines.

- **Fusion step:** A liquid obtained is transmitted in form of drops to machines connected to the furnace. These drops are blown to take the form of a mold provided in advance.

- **Shaping step**: This step is done automatically by machines to get the final form of the bottle.

- **Annealing step**: It allows lowering the tension of the final product, to get the bottles at the room temperature. After that, they are controlled and packaged.

### 4.2. Multi-criteria Evaluation of modeling techniques

#### 4.2.1. First experiment

**Alternatives**: The alternatives to compare are the five techniques for enterprise modeling (MERISE, GRAI, CIMOSA, PERA, GERAM).

**Criteria**: After considering the enterprise, we were able to identify ten criteria: functional, informational, resource and organizational views, formalism, cycle of life, function decision and flow decision, learning and ease of use.

**Performance table**: Taking into account these criteria, we obtain the performance table *(*Fig. 7*)*:

Fig.7 Performance table

**Classification of alternatives**: For any pair of action, we calculate the index of preference following equation (1), the result is grouped in a table called credibility matrix (Fig. 8).

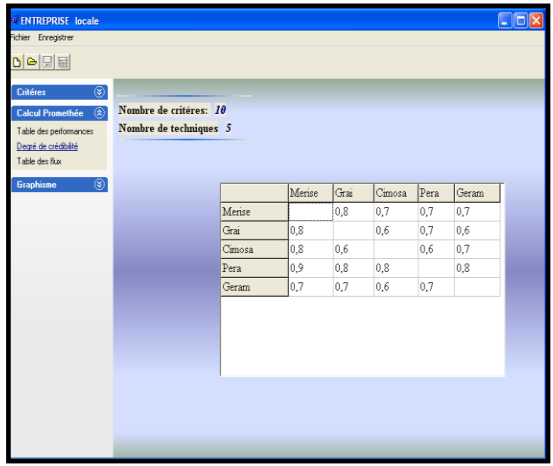

Fig. 8  Credibility matrix

For each action, we calculate the net flow according to equation (2) and we get the following ranking:

- Technical PERA with a net flow 0.074
- Technical CIMOSA with a net flow of 0
- Technical GERAM with a net flow of -0.012
- Technical GRAI with a net flow of -0.024.
- Technical MERISE with a net flow of -0.038.

The results of the classification techniques of enterprise modeling is given in the form of points reflecting the classification of five techniques for enterprise modeling, where the abscissas correspond to each technique and ordered to different values of net flows (Fig 9).

The most appropriate technique for enterprise modeling taking into consideration the criteria is PERA with a net flow of 0.074. The technique MERISE ranks last with a net flow of -0.038.

The result is justified because PERA defines all phases of the life cycle of an industrial entity (production of bottles) from its conceptualization to its implementation process. Moreover, it really takes into account the human aspect and position in architecture.

### 4.2.2. Second experiment

Our approach allows adding new technical enterprise modeling. Consider for example the GIM technique. The study of the GIM methodology allowed us to assign to each criterion.

The addition of GIM will be done by creating instances through the Protégé tool by adding the forums of GIM in the ontology. The application of the multi-criteria evaluation with the ten criteria identified for the enterprise and six technical enterprise modeling (MERISE, GRAI, CIMOSA, PERA, GERAM and GIM) has allowed us to obtain results of net flows and the graph as a histogram following Fig. 10.

- Technical PERA with a net flow of 0.078
- Technical GIM with a net flow 0.033
- Technical CIMOSA with a net flow of -0.022
- Technical GERAM with a net flow of -0.022
- Technical GRAI with a net flow of -0.022
- Technical MERISE with a net flow of -0.045.

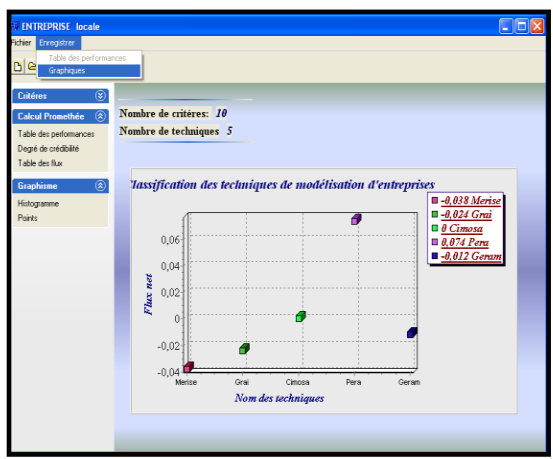

Fig. 9  Ranking of the five techniques
following ten criteria

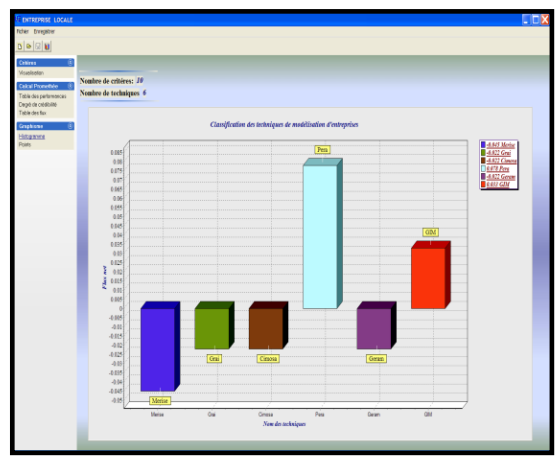

Fig.10 Ranking of the six techniques
following ten criteria

The technique PERA still ranks in first place with a net flow of 0.078. GIM ranks second with a net flow of 0.033. Note that CIMOSA, GERAM and GRAI are indifferent and ranked third with a net flow of -0.022. MERISE is still in last place with a net flow of -0045.

4.2.3. Third experiment

Resuming the previous two experiments by reducing the criteria taken into consideration. We remove the two decision criteria: the flow and function decisions.

**Cases of five alternatives**: The calculation of net flow of each of the five alternatives (MERISE, GRAI, CIMOSA, PERA, GERAM) is given as follows:

- Technical PERA with a net flow of 0.124
- Technical CIMOSA with a net inflow of 0.062
- Technical GERAM with a net inflow of 0.042.
- Technical MERISE with a net flow -0.063
- GRAI technique with a net flow of -0.167.

The technique PERA still ranks first with a net flow of 0.124, MERISE is the fourth with a net flow of -0063. GRAI takes the last position with a net flow of -0.067 as it is a technical based on aspects of the enterprise and the concerned and appropriate criteria were removed.

**Cases of six alternatives**: The ranking of the six modeling techniques enterprise (MERISE, GRAI, CIMOSA, PERA, GERAM and GIM) following the eight criteria is illustrated in the histogram in Fig.11.

- Technical CIMOSA with a net flow of +0.083
- Technical PERA with a net flow of +0.055
- Technical MERISE with a net flow of 0
- Technical GIM with a net flow of 0
- Technical GERAM with a net flow of -0.07
- Technical GRAI with a net flow of -0.07

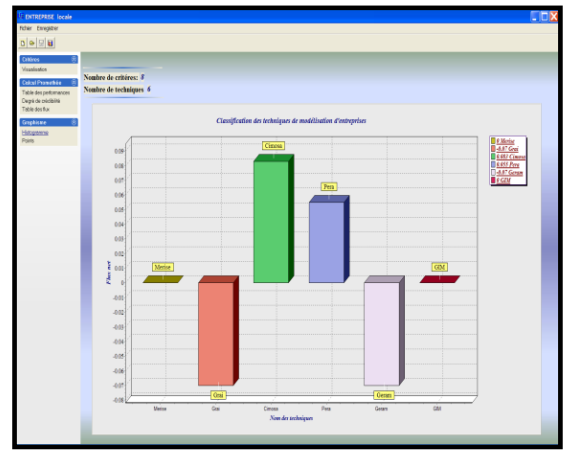

Fig 11. Ranking of the six technical following eight criteria

The technique CIMOSA ranks in first place with a net flow of 0.083. In the third position, MERISE and GIM are indifferent and last, the GERAM and GRAI techniques.

## 5. Conclusion

We have presented in this paper a domain ontology and multi-criteria analysis for enterprise modeling and these two views are supported in a meta-model.
The domain ontology allows representing techniques of enterprise modeling with their characteristics (criteria).
The multi-criteria system offers to patterns of Algerian enterprises an overview of modeling techniques.
To validate our proposal, the methodology is applied to a real enterprise dedicated to the manufacture of glass bottles.
We made many experiments showing clearly that the technique MERISE is not the most appropriate technique for efficient modeling.

So, the Algerian enterprise leaders must focus on other techniques more suitable to meet their expectations.
A dialog between the patterns who will express the needs of an enterprise and academic researchers who has knowledge of scientific techniques should be initiated.
At the end, these leaders may choose the most appropriate technique to close with their expressed requirements.
We believe that this approach represents a step toward an effective reorganization of the enterprise leading to development of industrial production enterprise.

[1]

**Sabria Hadj-Tayeb** received the Magister degree in computer science from the University of Sciences and Technology of Oran in 2009. Currently, she is an assistant teaching in the Preparatory school in Science and Technology of Oran (EPSTO) and she is a doctorate candidate. Her research focuses on two domains, the enterprise modeling and ontologies.

**Myriam Noureddine** is currently Assistant Professor in the department of computer science in the University of Sciences and Technology of Oran (USTO). Her research interests include the modeling methods for systems, especially in software engineering and manufacturing systems and also in the dependability area.